\documentclass{article}
 
\usepackage{authblk}
\usepackage{todonotes}
\usepackage{cite}
\usepackage{graphicx}
\usepackage{subcaption}
\usepackage[theorems]{tcolorbox}
\usepackage{ulem}
\usepackage[a4paper, total={6.5in, 9in}]{geometry}
\usepackage{url}

\graphicspath{{./figures/}}

\title{\vspace{-2ex} \Large \textbf{Mechanics Informed Fluoroscopy of Esophageal Transport}}

\author[1]{\normalsize Sourav Halder}
\author[2]{\normalsize Shashank Acharya}
\author[3]{\normalsize Wenjun Kou}
\author[3]{\normalsize Peter J. Kahrilas}
\author[3]{\normalsize John E. Pandolfino}
\author[1,2]{\normalsize Neelesh A. Patankar\thanks{Corresponding author: N.~A.~Patankar (\texttt{n-patankar@northwestern.edu})}}

\affil[1]{Theoretical and Applied Mechanics Program, McCormick School of Engineering,\newline
Northwestern University Technological Institute, 2145 Sheridan Road, Evanston, IL 60201 \vspace{1ex}}
\affil[2]{Department of Mechanical Engineering, McCormick School of Engineering, \newline
Northwestern University Technological Institute, 2145 Sheridan Road, Evanston, IL 60201 \vspace{1ex}}
\affil[3]{Division of Gastroenterology and Hepatology, Feinberg School of Medicine, \newline
Northwestern University, 676 North St. Clair Street, Arkes Suite 2330, Chicago, IL 60611 \vspace{1ex}}

\date{\vspace{-7ex}} % avoid printing date and remove the associated whitespace

\begin{document}
\maketitle

\begin{abstract}  

Fluoroscopy is a radiographic procedure for evaluating esophageal disorders such as achalasia, dysphasia and gastroesophageal reflux disease (GERD). It performs dynamic imaging of the swallowing process and provides anatomical detail and a qualitative idea of how well swallowed fluid is transported through the esophagus. In this work, we present a method called mechanics informed fluoroscopy (FluoroMech) that derives patient-specific quantitative information about esophageal function. FluoroMech uses a Convolutional Neural Network to perform segmentation of image sequences generated from the fluoroscopy, and the segmented images become input to a one-dimensional model that predicts the flow rate and pressure distribution in fluid transported through the esophagus. We have extended this model by developing a FluoroMech reference model to identify and estimate potential physiomarkers such as esophageal wall stiffness and active relaxation ahead of the peristaltic wave in the esophageal musculature. FluoroMech requires minimal computational time, and hence can potentially be applied clinically in the diagnosis of esophageal disorders.


\end{abstract}

\section{Introduction}
\label{intro}
The esophagus is a multi-layered muscular tube that transports food from the pharynx to the stomach with the help of neurally activated peristaltic contractions. Esophageal disorders can cause disruption of the process of swallowing, and a variety of symptoms including dysphagia, chest pain, and heartburn. Some of these disorders are gastroesophageal reflux disease (GERD), achalasia and eosinophilic esophagitis (EoE). The esophageal wall properties and neural activation play a significant role in esophageal transport and are therefore, potentially important physio-markers of esophageal disorders. Hence, understanding the biomechanics of esophageal transport can provide important insights into the nature of these disorders. The mechanics of esophageal transport have been studied extensively using analytical study \cite{li_brasseur_1993, Brasseur1987}, numerical simulations focused on particular aspects of esophageal transport \cite{brasseur1994, ghosh2005, yang2007}, and fully-resolved models which capture the interaction between a fluid bolus and the esophageal walls \cite{KOU2015, KOU2017}. 

\begin{figure*}[ht]
 \centering
 \includegraphics[scale = 0.4]{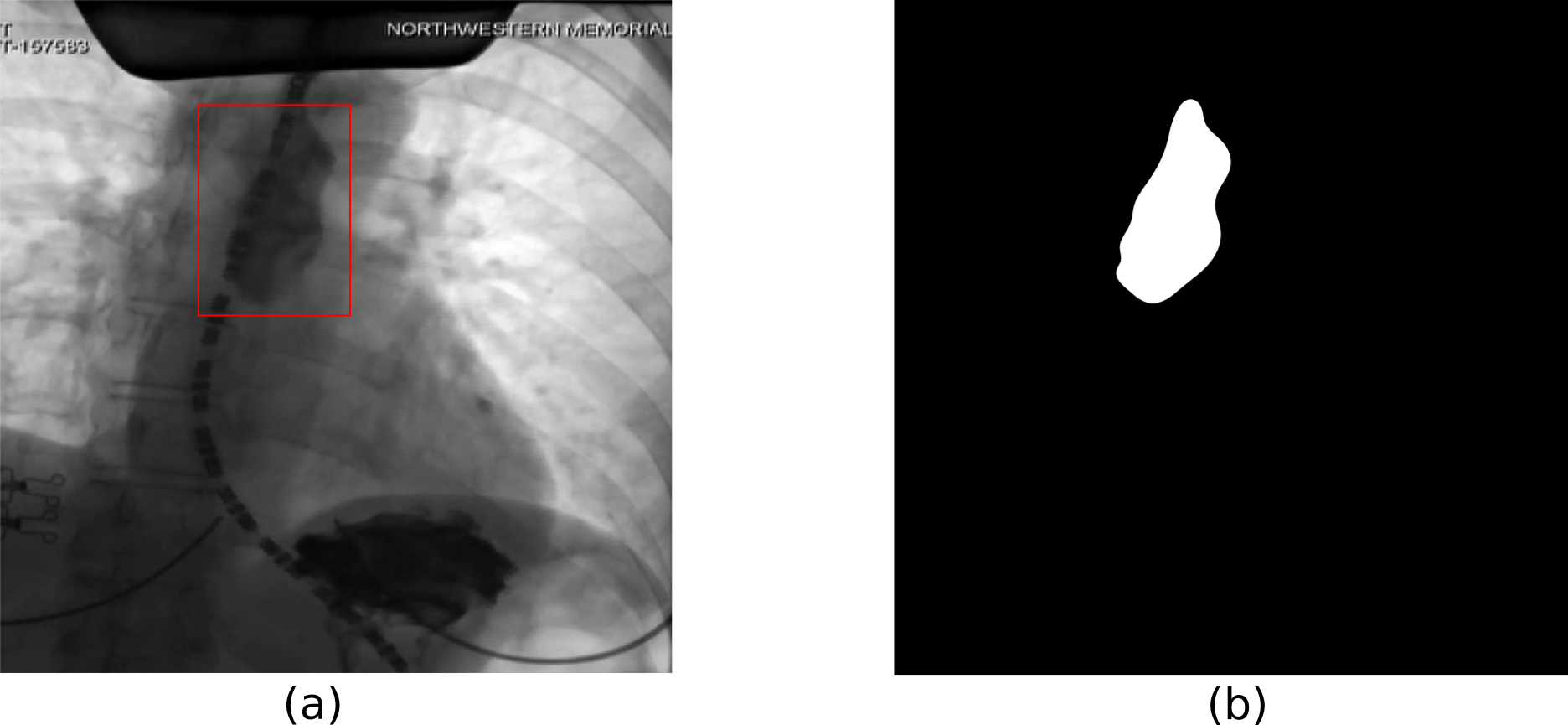} 
 \caption{Example of an image from VFSE performed jointly with HRM. (a) Original esophagram image. The bolus is the dark region inside the red box. The dashed curve is the HRM catheter passing through the esophagus to measure pressure along the length of the esophagus.  (b) Label image showing the bolus in white and the remainder as black}
 \label{fig-esophagram}
 \end{figure*}

Among the methods for evaluating esophageal disorders are the barium swallow esophagram, video fluoroscopy swallowing exam (VFSE), and high-resolution manometry (HRM). Both the esophagram and VFSE are radiographic tests that examine the dynamic function of the esophagus. In a barium swallow, barium is used as a contrast material to clearly delineate the esophageal lumen on an X-ray. This can reveal structural abnormalities of the esophagus and stomach such as hiatal hernia, diverticula, dilatation, etc. Video fluoroscopy uses the same concept, but creates a real-time X-ray movie of bolus transport. In HRM, a catheter is passed transnasally through the esophagus into the stomach (see Fig. 1(a)). The catheter incorporates pressure sensors along its length that quantify the intraluminal pressure along the length of the esophagus as the patient swallows fluid or food. HRM provides information about the strength and velocity of peristaltic contractions as well as the tone of the upper and lower esophageal sphincters. Barium swallow and VFSE are non-invasive but provide only qualitative information about esophageal transport. On the other hand, HRM is invasive, but provides precise quantitative information about esophageal contractility. In this work, we present a method to partially bridge the gap between these methods. We have developed a method called mechanics informed fluoroscopy (FluoroMech) that can be used along with VFSE to predict the flow rate, pressure and esophagus wall state, thereby providing quantitative information about bolus transport and esophageal contractility.

Previous studies have used data from fluoroscopy and manometry for the analysis with fluid mechanics \cite{brasseur1994, ghosh2005} and provided important insights into esophageal transport and mechanisms of a variety of disorders. However, a drawback of these techniques is that substantial time and effort is required to manually obtain the shape of the bolus from the fluoroscopy images and then perform an analysis based on that geometry. Since the geometry varies from patient to patient as well as for different swallows in the same patient, this entire analysis has to be repeated for every test sequence. Hence, these methods are not practical for clinical applications. FluoroMech uses deep learning to perform automatic segmentation of image sequences from fluoroscopy. This eliminates the tedious manual process of segmenting the fluoroscopy images, thereby making the process significantly faster and much more convenient. These segmented images delineate the outline of the bolus which then becomes input to a one-dimensional model that predicts the fluid flow rate and pressure. We also present a FluoroMech reference model that predicts the regional stiffness of the esophageal walls and the active relaxation at the locus of the bolus using the flow rate predicted from the one-dimensional model and the shape of the bolus. Our analysis requires minimal input from the user and requires minimal computational time. Hence, FluoroMech can be used in clinical applications, particularly to aid in making VFSE a more powerful non-invasive diagnosis tool.

%%%%%%%%%%%%%%%%%%%%%%%%%%%%%%%%%%%%%%%%%%%%%%%%%%%%%%%%%%%%%%%%%%%%%%%%%
\section{Image segmentation of fluoroscopy}
\label{part1_segmentation}
\begin{figure*}[ht]
 \centering
 \includegraphics[scale = 0.13]{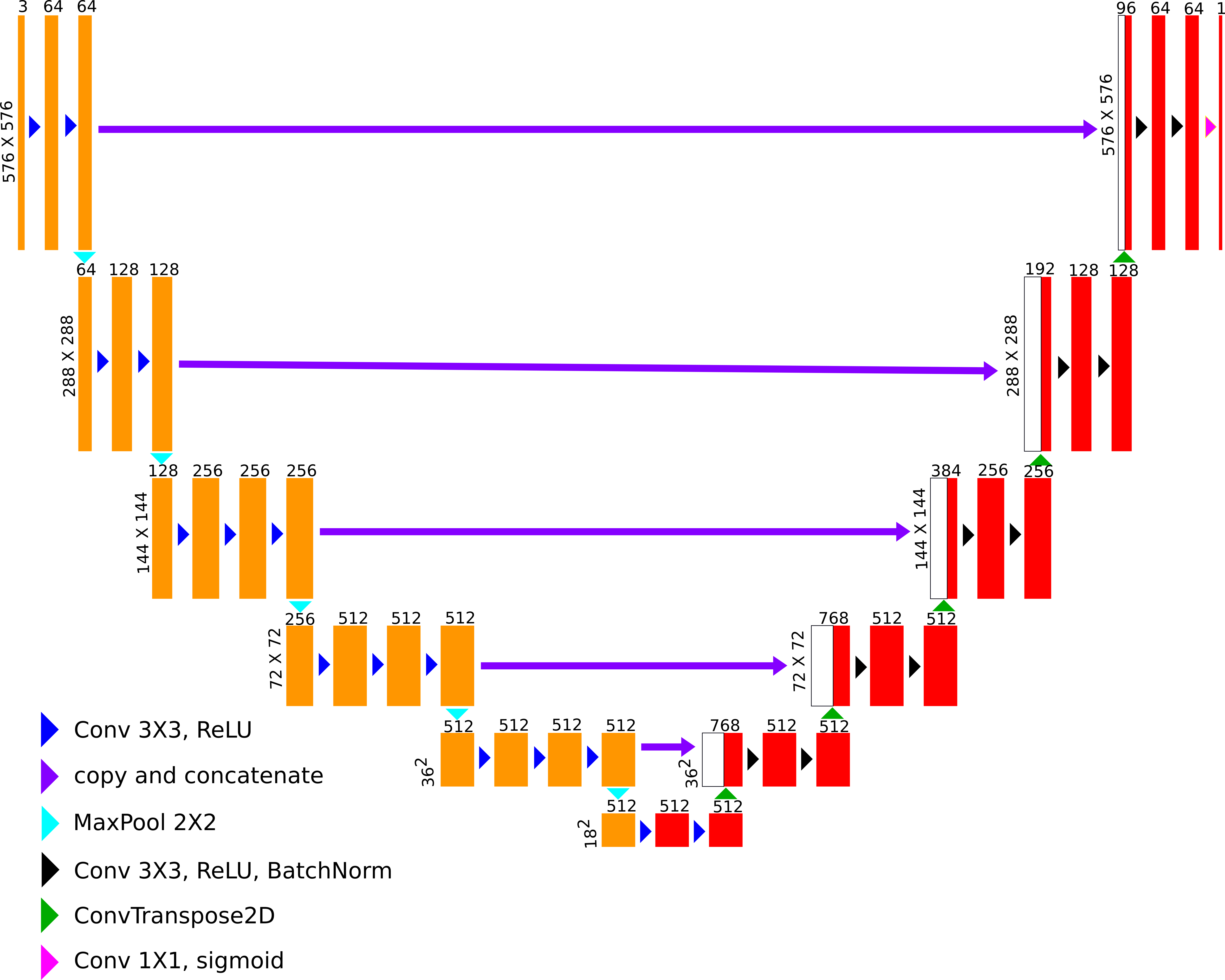} 
 \caption{Neural network architecture (based on TernausNet). The feature maps marked in yellow represents the VGG16 encoder pre-trained with the ImageNet dataset}
 \label{fig-nn_architecture}
 \end{figure*}
The volumetric quantification of fluid inside the esophagus can be approximated from two-dimensional images of the bolus in fluoroscopy. Fig. \ref{fig-esophagram}(a) shows an example of a single image from a sequence of images generated from a VFSE. In general, a single VFSE generates 100 – 500 images depending on the time taken for each swallow sequence to complete. Hence, it is not feasible to repeatatively manually outline the boundary of the barium bolus throughout the transport process. Rather, an automated technique is desirable to perform segmentation of the image sequences. There are several methods available in the literature for image segmentation such as thresholding \cite{sahoo1988}, region growing \cite{pal1993}, clustering \cite{coleman1979}, edge detection \cite{senthilkumar2018}, artificial neural networks \cite{ciresan2012, ronneberger2015, iglovikov2018, kayalibay2017, pham2000}. Surveys of the various image segmentation techniques used in medical applications were performed by \cite{pham2000, sharma2010}. 

\subsection{Neural network architecture}

In this work, we used a convolutional neural network architecture called TernausNet \cite{iglovikov2018} to perform image segmentation. TernausNet is a modified form of the classical UNet \cite{ronneberger2015} which consists of an encoder and decoder path with skip connections that combine feature maps from the encoder and decoder paths leading to precise localization. TernausNet takes advantage of transfer learning by replacing the encoder part of U-Net with VGG11/VGG16 network pretrained on ImageNet dataset, which consists of millions of images. Therefore, the low level features learned from a huge dataset can be efficiently utilized and the total number of parameters to be learned is reduced significantly. In this work, the encoder part consists of VGG16. The decoder path is similar to that of the original TernausNet with the slight modification of having two sets of Conv 3X3 and ReLU at each level instead of one. The full network architecture is shown in Fig. \ref{fig-nn_architecture}. In order to prevent over-fitting, we have also introduced batch normalization after every convolution layer in the decoder section. The whole network consists of 36,319,201 parameters, of which 28,676,001 were pre-trained and 7,643,200 were trained.

\subsection{Dataset}

\begin{figure*}[ht]
 \centering
 \includegraphics[scale = 0.2]{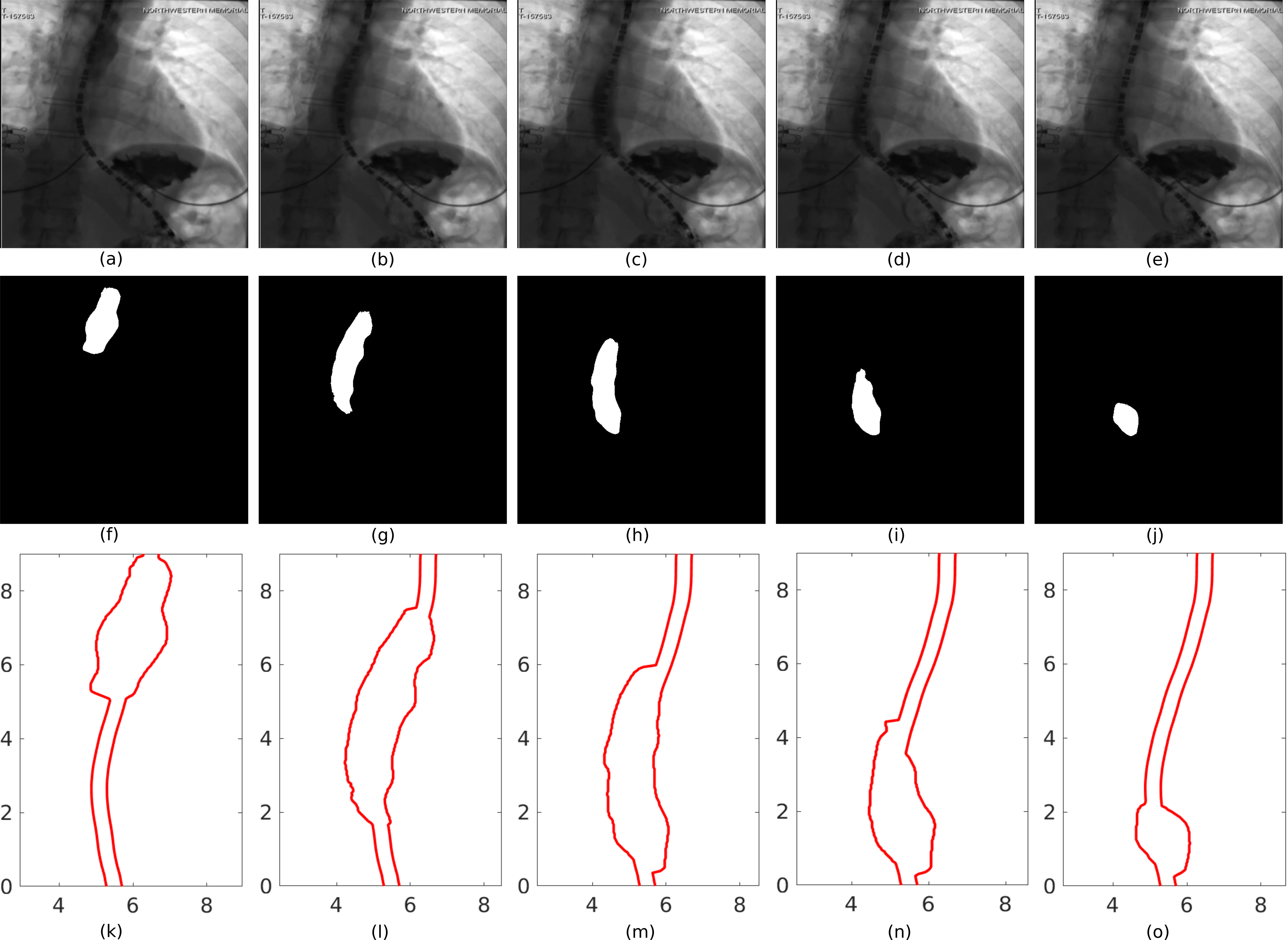} 
 \caption{Segmentation of image frames of a VFSE. (a)-(e) Bolus transported from the proximal to the distal end of the esophagus and emptying in to the stomach, (f)-(j) corresponding image segmentation, (k)-(o) corresponding outline of the esophageal lumen for analysis}
 \label{fig-segmentation}
\end{figure*}

We used 136 esophagram images from 99 swallows (from 14 different patients) showing different instants of the transport process. Each of these images has a size of $576 \times 576$ pixels. We obtained these images from VFSE done in association with HRM recordings (hence, the presence of the catheter in Fig. \ref{fig-esophagram}(a)). These images were manually segmented for labeling as shown in Fig. \ref{fig-esophagram}(b). Since the training dataset is very small, it makes sense to take advantage of transfer learning through the pre-trained encoder in order to prevent over-fitting. In addition to that, we have also implemented various image augmentations such as rotation, height and width shifts, varying brightness, shearing, piecewise affine and scaling. This is done to prevent overfitting due to the lack of a huge dataset and to introduce generalizability into the model. These augmentations were applied randomly to varying extents, the range of which is provided in Table \ref{tab-img_augmentation}.
\begin{table}[ht]
\centering
 	\caption{Details of data augmentations}
	\begin{tabular}{l | l l}
		\hline\noalign{\smallskip}
 		\textbf{Augmentation type} & \textbf{Range}  \\
  			& min & max \\
 		\noalign{\smallskip}\hline\noalign{\smallskip}
 		Rotation & $-10.0^{\mathrm{o}}$ & $10.0^{\mathrm{o}}$\\
 		Width shift & $-10\%$ & $10\%$\\ 
 		Height shift & $-10\%$ & $10\%$\\
 		Brightness & $50\%$ & $150\%$\\
 		Shear & $-5^{\mathrm{o}}$ & $5^{\mathrm{o}}$\\
 		Zoom & $80\%$ & $120\%$\\
 		Piecewise affine & $0$ & $0.03$\\
 		\noalign{\smallskip}\hline
	\end{tabular}
 	\label{tab-img_augmentation}
\end{table}

\subsection{Training and segmentation prediction}
The dataset of 136 images was divided into two parts: 112 images for training and 24 images for validation. This is a semantic segmentation problem, wherein each pixel belongs to one of the two classes: 1 for bolus and 0 for the background. We used a sum of binary crossentropy ($BCE$) and negative Intersection over Union ($IOU$) loss functions as the total loss ($L$) defined as follows:
\begin{align}
 BCE &= -\frac{1}{N}\sum\limits_{i=1}^N\left[y_i\mathrm{log}\left(\hat{y_i}\right) + \left(1-		y_i\right)\mathrm{log}\left(1-\hat{y_i}\right)\right], \\
 IOU &= \frac{1}{N}\sum\limits_{i=1}^N \frac{y_i\hat{y_i}+\epsilon}{y_i+\hat{y_i} - y_i\hat{y_i} + \epsilon}, \\
 L &= BCE - IOU,
\end{align}
 where $N$ is the total number of pixels in the output, $y_i$ and $\hat{y_i}$ are the target binary value and predicted value of the $i$-th pixel respectively. The $\epsilon$ added in the numerator and denominator of $IOU$ is a small number ($=10^{-7}$), which is introduced to calculate $IOU$ over both the classes: bolus and background. In order to evaluate the performance of the model, the predicted images were converted to binary form using various thresholds between 0.5 and 1.0, and $IOU$ were calculated for each of them and then averaged.
 
The model was trained for 200 epochs with batches of 2 images using Keras, a high-level neural networks API \cite{chollet2015keras}, which runs on top of TensorFlow \cite{tensorflow2015-whitepaper}, to train the network. The training was performed using RMSProp optimization algorithm with a learning rate of 0.001. The Intersection over Union for the validation set obtained at the end of the training was 0.75. The segmented output images were converted to binary form using a threshold of 0.5 for the final output. Some of the image frames for a sequence of images generated from a VFSE and the predicted segmentation of those images after thresholding are shown in Fig. \ref{fig-segmentation}(a)-(j).

\subsection{Post-processing}

The sharp interface between the white and dark regions of the segmented images marks the outline of the bolus. It gives the shape of the inner mucosal surface of the esophagus at the location of the bolus, but no information about regions of esophageal contraction or relaxation. The diameter of the catheter (dashed curve in Fig. \ref{fig-esophagram}(a)) is approximately 4.2 mm. We use this as the scale for mapping the pixel data to length. In some image frames, the relaxed diameter of the esophagus can be identified at some locations along the length due to remnant barium lining the lumen. In order to simplify our analysis, we assume that this diameter is the relaxed diameter of the esophagus throughout its length, although, in reality, the esophagus may be collapsed or the inner diameter may vary along the length \cite{xia2009}. Our analysis shows that this assumption does not significantly affect the calculation of intra-bolus pressure. Fig. \ref{fig-segmentation}(k-o) displays the shape of the esophageal lumen which is used for the analysis described in the next section. The semantic segmentation performed on the esophagram images basically assigns each pixel to one of the two classes: bolus (white region) and the remainder (dark region). The resulting segmented images do not show a smooth boundary for the bolus and is irregular at the scale of the resolution of the original image. Also, since the segmentation is done on each of the images separately, the continuity between the images at consecutive time frames is broken. Therefore, the pixel data were smoothed both in space and time without the loss of bolus geometry detail. The smoothing is performed by Gaussian weighted moving average over a window of 10 and 30 points in space and time respectively.

%%%%%%%%%%%%%%%%%%%%%%%%%%%%%%%%%%%%%%%%%%%%%%%%%%%%%%%%%%%%%%%

\section{Mathematical formulation}
\label{part2_math_formulation}

\subsection{Governing equations}

An important aspect of patient-specific analysis of esophageal transport is obtaining the flow rate and pressure field inside the esophagus with reasonable accuracy using limited computational resources and time. To that extent, we have used the formulation of a one-dimensional flow through a flexible tube \cite{Barnard1966, ottesen2003, manopoulos2006, kamm_shapiro_1979} in FluoroMech to model the transport process. The mass and momentum conservation equations in one-dimension are follows:
\begin{align}
 \frac{\partial A}{\partial t} + \frac{\partial Q}{\partial x} &=0, \label{eqn-continuity} \\
 \frac{\partial Q}{\partial t} + \frac{\partial}{\partial x}\left(\frac{4}{3}\frac{Q^2}{A}\right)+\frac{A}{\rho}\frac{\partial P}{\partial x} + \frac{8\pi\mu Q}{\rho A} &= 0, \label{eqn-momentum}
\end{align}
where $\rho$ is the density of the fluid, $\mu$ is viscosity of the fluid, $A$ is the cross-sectional area of the esophagus, $Q$ is the flow rate, $P$ is the pressure, $t$ is the time and $x$ is the spatial coordinate along the length of the esophaus with its positive direction defined as moving from the pharynx to the stamach. We assume that the cross-section of the esophagus, $A(x,t)$ is elliptical in shape, with the major axis in the plane shown in the esophagram. The flow rate, $Q(x,t)$ is defined as follows:
\begin{align}
Q=u_{m}A,
\end{align}
where $u_{m}$ is the mean velocity of the fluid across a cross-section. The factor $4/3$ in Eq. \ref{eqn-momentum} comes from assuming a parabolic velocity profile perpendicular to the direction of flow.

Eqs. \ref{eqn-continuity} and \ref{eqn-momentum} are non-dimensionalized to the following form:
\begin{align}
 \frac{\partial \alpha}{\partial \tau} + \frac{\partial q}{\partial \chi} =0, \label{eqn-continuity_nd} \\
 \frac{\partial q}{\partial \tau} + \frac{\partial}{\partial \chi}\left(\frac{4}{3}\frac{q^2}{\alpha}\right)+\alpha\frac{\partial p}{\partial \chi} + \psi\frac{q}{\alpha} = 0, \label{eqn-momentum_nd}
\end{align}
where, $\chi=x/L$, $\alpha = A/A_o$, $p=P/(\rho c^2)$, $q=Q/(A_oc)$, $\tau=ct/L$ and $\psi=8\pi\mu L/(\rho c A_o) $. Here $L$ is the length of the esophagus visible in the esophagram, $A_o$ is the relaxed cross-sectional area of the esophageal lumen, and $c$ is the average velocity of the center of the bolus. The center of the bolus ($x_b$) was located at each time instant using the following relation:
\begin{align}
x_b = \frac{\int_0^L x(A-A_o)dx}{\int_0^L (A-A_o)dx}  .   \label{eqn-bolus_center}
\end{align} 

In this instance, $L=11.86$ cm, $A_o=59.04$ $\textrm{mm}^2$ and $c=3.5$ cm/s. We assumed the properties of water (at STP) for the the swallowed fluid, i.e $\rho = 1000$ $\textrm{kg/m}^3$, $\mu=8.9\times10^{-4}$ Pa. s. Using these values, we get $\psi=2.413$. The total time required for bolus to be transported through the esophagus was 5.1 seconds.

\subsection{Initial and Boundary Conditions}
\begin{figure*}[h]
 \centering
 \includegraphics[scale = 0.3]{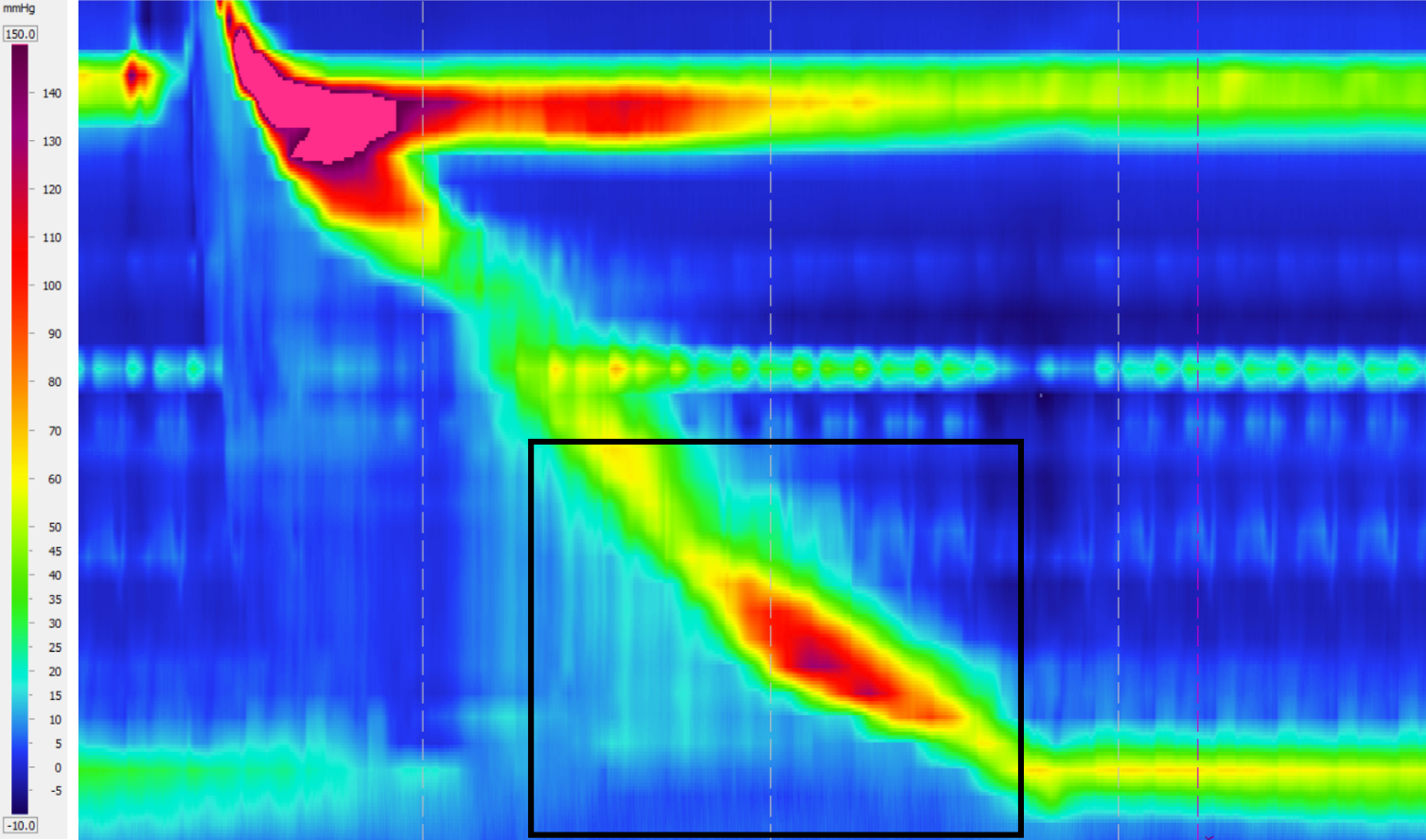} 
 \caption{Esophageal pressure topography generated from pressure sensors on the HRM catheter. The horizontal axis represents time and the vertical axis represents the length along the esophagus. The rectangular box shows the location of the EPT corresponding to the fluoroscopy}
 \label{fig-ept}
\end{figure*}
The boundary conditions imposed on the FluoroMech model depend on the behavior of the upper and lower esophageal sphincters at the proximal and distal ends of the esophagus respectively. The upper esophageal sphincter (UES) is located at the distal end of the pharynx and remains closed \cite{lang1997}  in order to prevent the entry of air into the esophagus during breathing and reflux of the bolus from the esophagus back into the pharynx \cite{mittal2011}. It relaxes for 0.32 - 0.5 seconds \cite{jacob1989} in order to allow the bolus to enter the esophagus. The esophageal pressure topography in Fig. \ref{fig-ept} illustrates this behavior of the UES. The horizontal high-pressure zone at the top marks the location of the UES, which remains contracted on the HRM catheter. It opens only to allow the bolus to enter the esophagus, which is visible as the break in the continuous high-pressure zone due to relaxation. The oblique band of pressure represents the peristaltic contraction which propels the bolus along the esophagus. Hence, the location of the proximal end of the bolus can be roughly identified to be just distal to the contraction. The lower esophageal sphincter (LES) is marked by the lower horizontal high-pressure zone. There is a break in high pressure in this location soon after the bolus enters the esophagus. This represents the relaxation of the LES to facilitate the bolus emptying into the stomach.
In our analysis, the bolus is already inside the esophagus, and so the UES is closed. Hence, there is no flow at the entry, i.e. $q(\chi=0,\tau)=0$. Additionally, we assume that there was no initial flow inside the esophagus, i.e. $q(\chi,\tau=0)=0$. Since Eq. \ref{eqn-momentum_nd}, contains a first-order derivative form for pressure, we have specified a reference value for the pressure ($p$) at the distal end equal to the intragastric pressure which in this senario is 21.4 mmHg. This is a reasonable assumption because the LES relaxes immediately at the beginning of a swallow, and the distal end of the esophagus experiences the intragastric pressure.

\subsection{Enforcing volume conservation}
\begin{figure*}[h]
\captionsetup[subfigure]{justification=centering}
\begin{subfigure}[c]{.45\textwidth}
  \centering
  \includegraphics[scale=0.55]{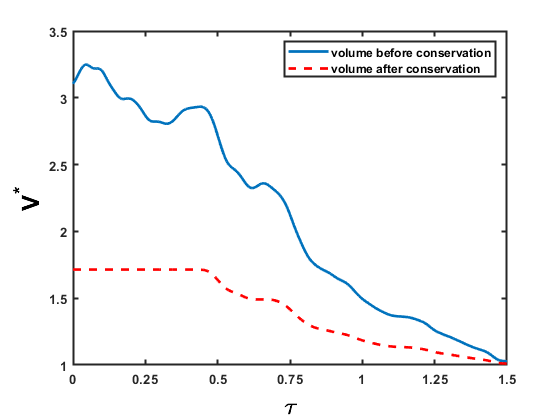}
  \caption{Volume inside the esophagus}
  \label{fig-vol}
\end{subfigure}
\begin{subfigure}[c]{.49\textwidth}
  \centering
  \includegraphics[scale=0.25]{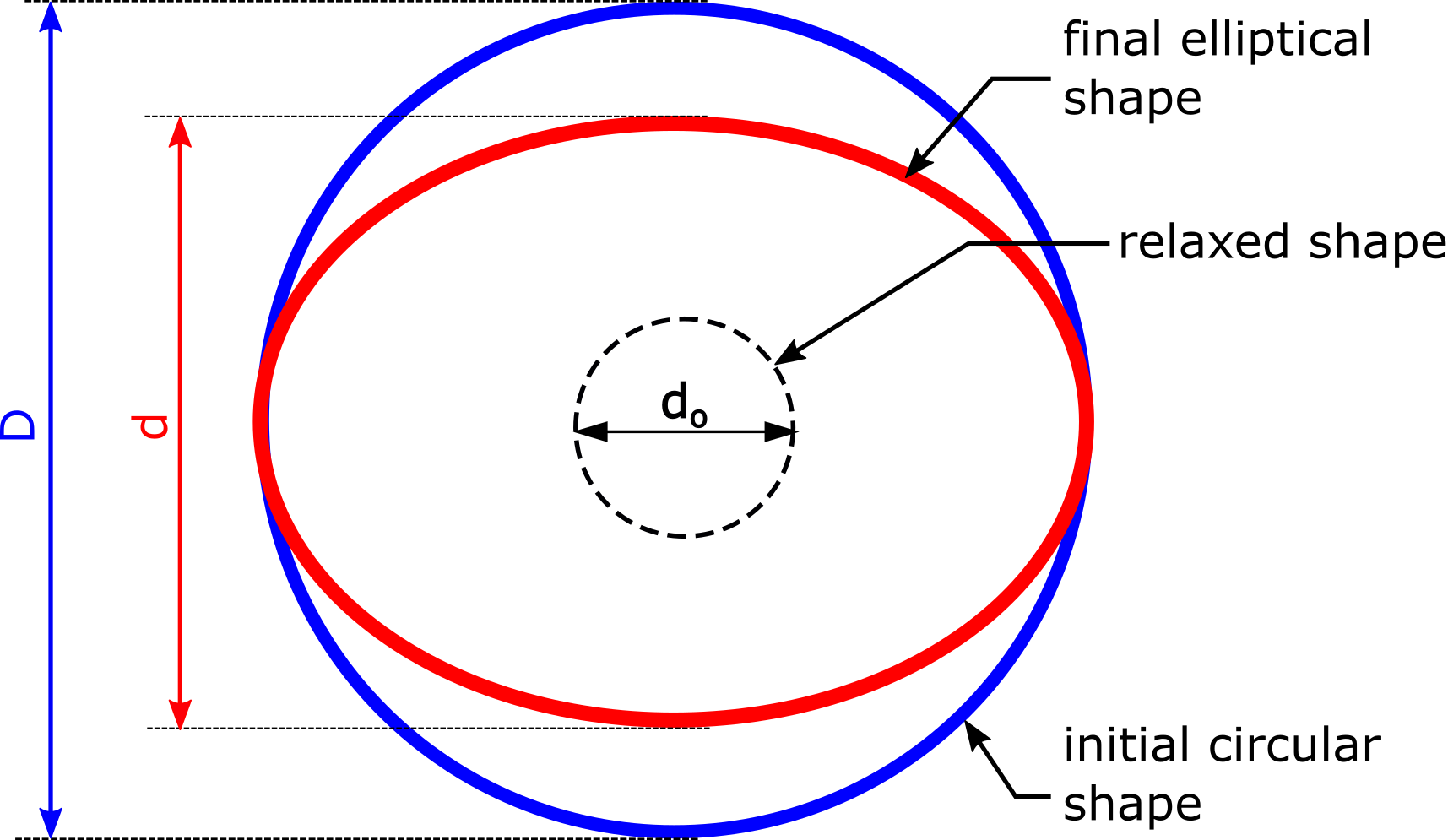}
  \caption{Schematic diagram of shape modification for volume conservation}
  \label{fig-shape}
\end{subfigure}
\begin{subfigure}{1.0\textwidth}
  \centering
  \includegraphics[scale=0.65]{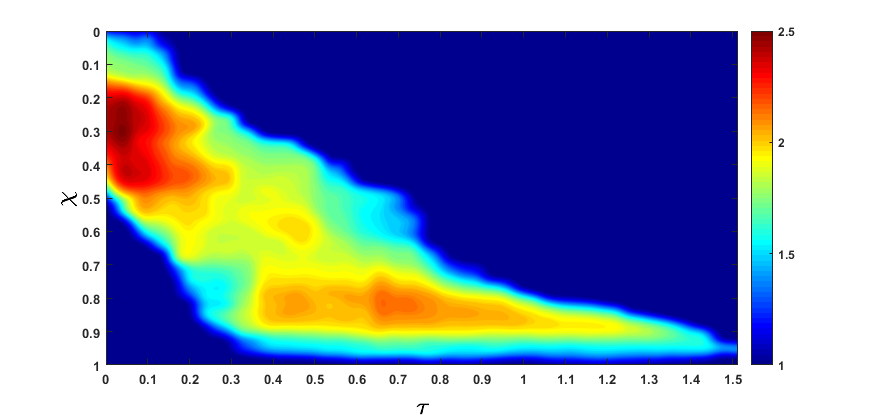}
  \caption{Ratio of major and minor axis (D/d).}
  \label{fig-dia_ratio}
\end{subfigure}
\caption{Enforcing volume conservation}
\label{fig-volume}
\end{figure*}
The fluoroscopy images show only a two-dimensional projection of the esophagus. Assuming a circular cross-section, we calculated the total volume of fluid inside the esophagus (as shown in Fig. \ref{fig-vol}). The volume $V$ is non-dimensionalized using the product of the non-distended cross-sectional area and the length of the esophagus, i.e. $V^* = V/(A_oL)$. In terms of the volume of fluid inside the esophagus, the bolus transport is categorized into two parts: pure transport (no flow at $\chi=1$) and emptying. From the VFSE image sequences we have observed that the transport without emptying occurs untill $\tau=0.5$, and then the volume within the esophagus decreases continuously. Since we have no flow boundary condition at the proximal end, the volume within the esophagus can never exceed the total volume at $\tau=0$. However, during pure transport at $\tau<0.5$, there are some fluctuations in the calculated volume within the esophagus. This can be attributed to our calculation of volume assuming the esophagus is perfectly circular in cross-section at all times. Since we have no information about the actual shape of the cross-section at $\tau =0$, we can neither use the calculated volume at $\tau=0$, nor the maximum calculated volume during the whole transport in order to enforce volume conservation. 

In reality, the shape of the esophagus cross-section is elliptical \cite{xia2009}, with the major axis being observed in the fluoroscopy images. The volume of fluid swallowed for every test ($V_o$) is 5 mL. Using this information, we scaled the circular cross-sectional area to an elliptical shape (see Fig. \ref{fig-shape}) so that the total volume inside the bolus is 5 mL. The scaling is performed in the following manner:
\begin{align}
A^* = A_o + \beta\left(A-A_o\right); \quad \quad \beta = \frac{V_o}{\int_0^L \left(A-A_o\right)dx}, \label{eqn-volume_conserve}
\end{align}
where $A^*$ is the scaled cross-sectional area to conserve volume and $\beta$ is the scaling factor. This method scales only the cross-section of the esophagus at the bolus location and does not change the relaxed sections.

We enforced a constant volume during pure transport (as shown by the red dashed line in Fig. \ref{fig-vol}) untill $\tau=0.5$.  However, during emptying ($\tau>0.5$), the volume within the esophagus begins to decrease, so, we cannot scale the volume using a reference value. Therefore, the volume is scaled using $\beta$ calculated at the beginning of the emptying process ($\tau=0.5$). In general, the shape of the esophageal cross-section varies along its length, and the shape it takes when distended depends upon the material properties of the wall. The $\beta$ calculated at each time step during pure transport gives a measure of the shape of the cross-section. As the bolus is transported through the esophagus, the $\beta$ takes on different values, thereby estimating the shape in finite segments along the length. At the beginning of emptying, the distal end of the bolus has already reached the end of the esophagus. After this, the length of the bolus progressively decreases without moving any forward. Therefore, the $\beta$ calculated at $\tau=0.5$ is a reasonable scaling for volume during emptying. 

During emptying, the volume within the esophagus must decrease. Therefore, if at any instant, the calculated volume inside the bolus is greater than it was in the previous time instant, we force the volume at the current instant to be equal to that of the previous instant. The effect of the volume correction on the diameter of the esophagus is shown in Fig. \ref{fig-vol}.  The ratio of the major and minor diameter of the scaled elliptical shape of the cross-section is shown in Fig. \ref{fig-dia_ratio}. The diameters are non-dimensionalized using the relaxed diameter of the esophagus.  We see that the change in diameter occurs only at the location of the bolus since the ratio of major to minor diameter remains equal to 1 for the remainder of the esophagus. Comparing Fig. \ref{fig-vol} and \ref{fig-dia_ratio}, we see that the maximum changes in diameter occur before $\tau=0.2$ when the difference between the reference volume and the calculated volume is maximal,  scaling the cross-section into a flatter ellipse.  

\subsection{Numerical solution}
\begin{figure*}[h]
 \centering
 \includegraphics[scale = 0.4]{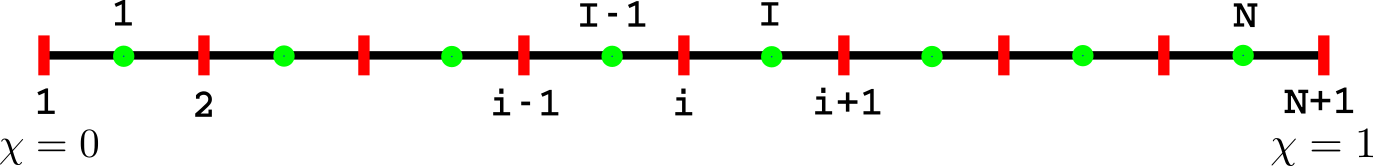} 
 \caption{Staggered meshing of the domain. The cell boundaries and centers are shown in red and green respectively}
 \label{fig-domain}
\end{figure*}
Using the cross-sectional areas $\alpha$ obtained after enforcing volume conservation, we solved for $q$ and $p$ in Eqs. \ref{eqn-continuity_nd} and \ref{eqn-momentum_nd} using the finite volume method. The flow rate ($q$) is calculated by solving Eq. \ref{eqn-continuity_nd}. A staggered grid was used to discretize the domain as shown in Fig. \ref{fig-domain}. The flow rate $q$ is calculated at the cell boundaries and pressure $p$ is calculated at the cell centers. The cross-sectional area $\alpha$ is known for both the cell boundaries and centers. The quantities specified at the cell centers have subscripts in upper case, and those at the cell boundaries in lower case. The superscript $o$ represent the value of a quantity in the previous time instant. Eq. \ref{eqn-continuity_nd} is solved using a fully-implicit method with the following discretized form:
\begin{align}
q_i = q_{i-1} + \frac{\Delta \chi}{\Delta \tau}\left(\alpha_{I-1} - \alpha^{o}_{I-1}\right),
\end{align}
where, $N$ is the total number of cells, $i, I = 2 , 3, ..., N, (N+1)$. Using the calculated values of $q$ and the known values of $\alpha$, we calculated the values of $p$ at the cell centers using the following discretized form:
\begin{align}
p_I = &p_{I+1} + \left(\frac{\Delta \chi}{\alpha_i}\right)\frac{q_{i+1}-q_{i+1}^o}{\Delta \tau} + \psi\frac{q_{i+1}\Delta \chi}{\alpha_{i+1}^2} \nonumber \\
& - \frac{1}{3\alpha_{i+1}}\left[\frac{(q_{i+2}+q_{i+1})^2}{\alpha_{I+1}} - \frac{(q_{i}+q_{i+1})^2}{\alpha_{I}}\right],
\end{align}
where $i, I = 1, 2, ..., (N-1)$. In this simulation, the total number of time steps and the total number of cells used were 510 and 171 respectively. Using these values, $\Delta \tau$ and $\Delta \chi$ was calculated as 0.003 and 0.006 respectively. The above mentioned numerical solution was implemented using MATLAB ver. R2018b. 

\subsection{Pressure variation with the shape and speed of the bolus}
\label{sec-pressure_variation}

\begin{figure*}[h]
 \centering
 \includegraphics[scale = 0.18]{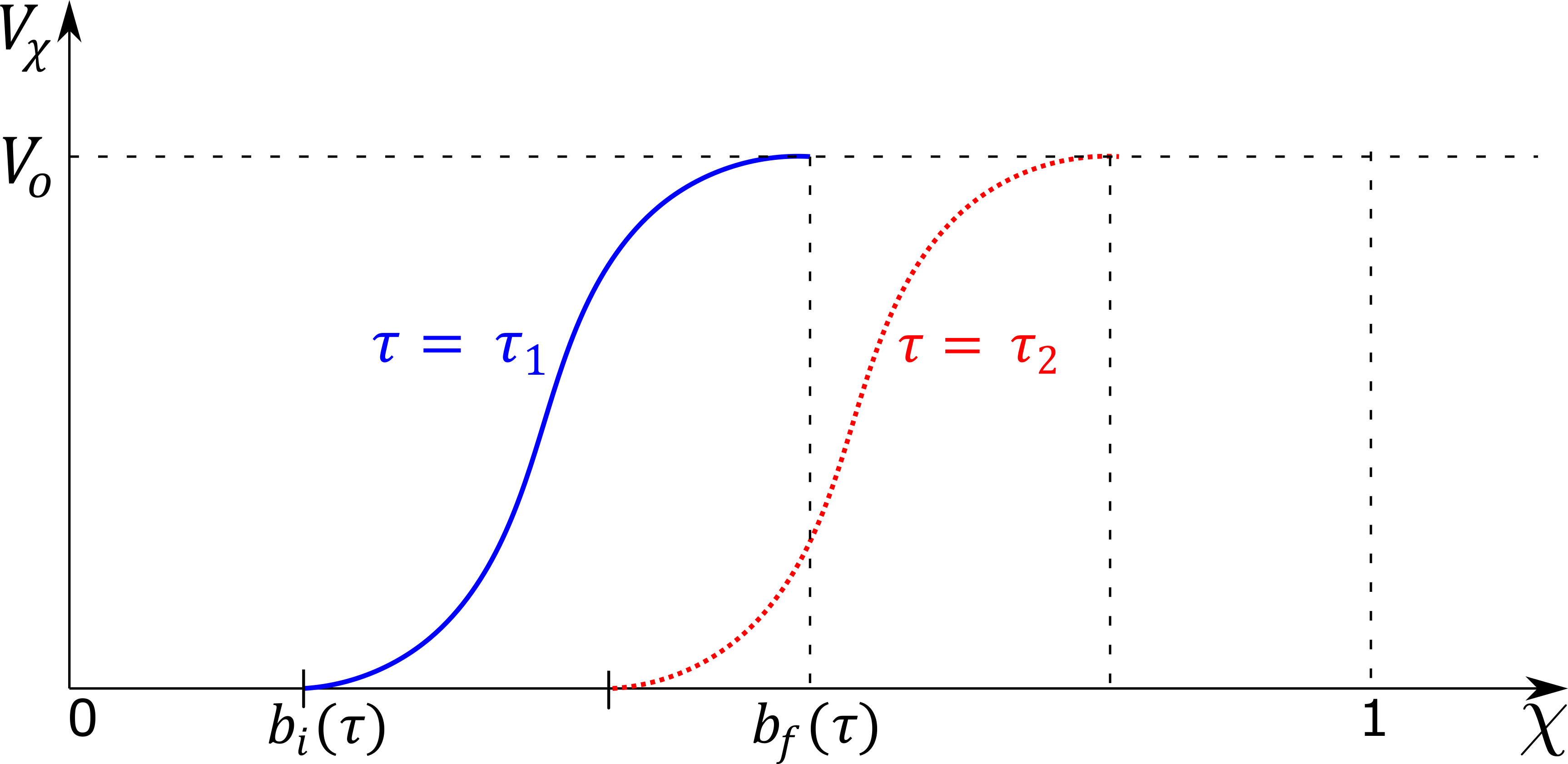} 
 \caption{Volume inside the esophagus upto $\chi$}
 \label{fig-volumeX}
\end{figure*}
The intrabolus pressure during pure transport (before emptying) can be used to calculate the esophageal wall stiffness and active relaxation. Therefore, it is necessary to identify and estimate the factors that lead to pressure variations.
The flow rate is calculated by integrating Eq. \ref{eqn-continuity_nd} with respect to $\chi$ as follows:
\begin{align}
q=-\frac{\partial}{\partial\tau}\int_0^{\chi}\alpha d\chi =-\frac{\partial}{\partial\tau}\int_0^{\chi}\left(\alpha'+1\right) d\chi= -\frac{\partial V_{\chi}}{\partial \tau}, \label{eqn-flowrate_analytic}
\end{align}
whereby the cross-sectional area $\alpha$ is decomposed into the non-distended cross-section area (which is equal to 1 in non-dimensional form) and the extra volume present only inside the bolus ($\alpha'$). $V_{\chi}$ is the volume inside the bolus calculated by integrating $\alpha'$ upto $\chi$. The variation of $V_{\chi}$ is shown in Fig. \ref{fig-volumeX}, where $b_i$ and $b_f$ represent the location of the proximal and distal end of the bolus respectively. Since the total volume $V_o$ is conserved within the esophagus prior to the start of emptying, $V_{\chi}=0$ for $\chi<b_i$ and $V_{\chi}=V_o$ for $\chi>b_f$. Therefore, using Eq. \ref{eqn-flowrate_analytic}, we get $q=0$ for $\chi<b_i$ and $\chi>b_f$. 

The effect of fluid viscosity is captured by the term $\psi\frac{q}{\alpha}$ in Eq. \ref{eqn-momentum_nd}. For a fluid with viscosity similar to water, we have observed that the viscous term is negligible compared to the other terms of Eq. \ref{eqn-momentum_nd}. Therefore, we assume the flow to be inviscid and with a flat velocity profile, and write Eq. \ref{eqn-momentum_nd} as follows:
\begin{align}
\frac{\partial}{\partial \tau}\left(\frac{q}{\alpha}\right) + \frac{\partial}{\partial \chi}\left[\frac{1}{2}\left(\frac{q}{\alpha}\right)^2\right]+\frac{\partial p}{\partial \chi}= 0. \label{eqn-momentumRM}
\end{align}
On integrating Eq. \ref{eqn-momentumRM} with respect to $\chi$ from the distal end and using Eq. \ref{eqn-flowrate_analytic}, we get
\begin{align}
p = p_1+\frac{1}{2\alpha_1^2}q_1^2-\frac{1}{2\alpha^2}\left(\frac{\partial V_{\chi}}{\partial \tau}\right)^2 -\frac{\partial}{\partial \tau}\int_1^{\chi} \frac{1}{\alpha}\frac{\partial V_{\chi}}{\partial \tau}d\chi, \label{eqn-p_x}
\end{align}
where the subscript $\chi$ indicates the location at which each of the quantities are calculated, and $p_1$, $q_1$ and $\alpha_1$ are the pressure, flow rate and cross-sectional area at $\chi=1$ respectively. Before emptying begins, $q_1=0$ so, the second term of Eq. \ref{eqn-p_x} becomes equal to 0. Since $q=0$ for $\chi<b_i$ and $\chi>b_f$, Eq. \ref{eqn-momentumRM} implies $\partial p/\partial\chi=0$ at these points. Therefore, $p=p_1$ for all $\chi>b_f$ and $p=p_i$ for all $\chi<b_i$ wherein $p_i$ is the pressure at $\chi=b_i$. Our results showed that the maximum and minimum values of $p$ are observed at $\chi=b_i$, so $p_i$ was used as the upper and lower bound of the pressure variation for each time step. At $\chi=b_i$, $V_{\chi}=0$ so, the third term of Eq. \ref{eqn-p_x} becomes equal to 0. On applying these arguments to Eq. \ref{eqn-p_x}, we get
\begin{align}
p_i =p_1 -\frac{\partial}{\partial \tau}\int_{b_i}^{b_f} \frac{q}{\alpha}d\chi. \label{eqn-p_i}
\end{align} 

We define an average estimate of the bulk velocity of the bolus, $u_b$ for every time instant as follows
\begin{align}
u_b=\frac{1}{L_b}\int_{b_i}^{b_f}\frac{q}{\alpha} d\chi, \label{eqn-u_b}
\end{align}
where, $L_b = b_f-b_i$, is the length of the bolus at each instant of time. Using Eq. \ref{eqn-p_i} and \ref{eqn-u_b}, we get
\begin{align}
p_i =p_1-u_b\frac{\partial L_b}{\partial \tau} - L_b\frac{\partial u_b}{\partial \tau}. \label{eqn-pressure_fluc}
\end{align}
From Eq. \ref{eqn-pressure_fluc}, we see that the pressure variation between its maximum and minimum can be attributed to the variation of the bulk velocity of the bolus and the change in length of the bolus during transport through the esophagus.

\subsection{Estimating the stiffness and relaxation of esophageal wall}
\label{sec-stiffness}
\begin{figure*}[h]
 \centering
 \includegraphics[scale = 0.8]{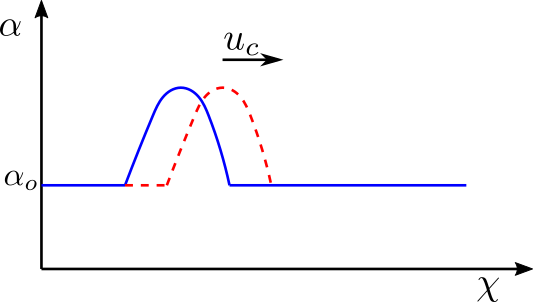} 
 \caption{Bolus transport in the FluoroMech reference model}
 \label{fig-reference_model}
\end{figure*}
\begin{figure*}[h]
\captionsetup[subfigure]{justification=centering}
\begin{subfigure}[c]{.49\textwidth}
  \centering
  \includegraphics[scale=0.2]{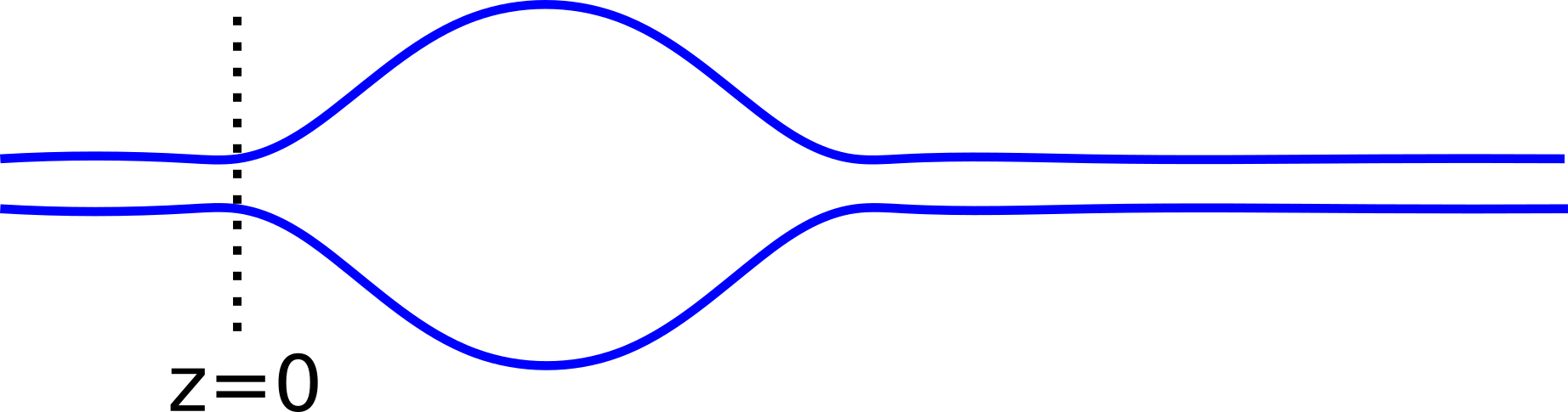}
  \caption{Bolus shape with relaxation}
  \label{fig-with_relax}
\end{subfigure}
\begin{subfigure}[c]{.49\textwidth}
  \centering
  \includegraphics[scale=0.2]{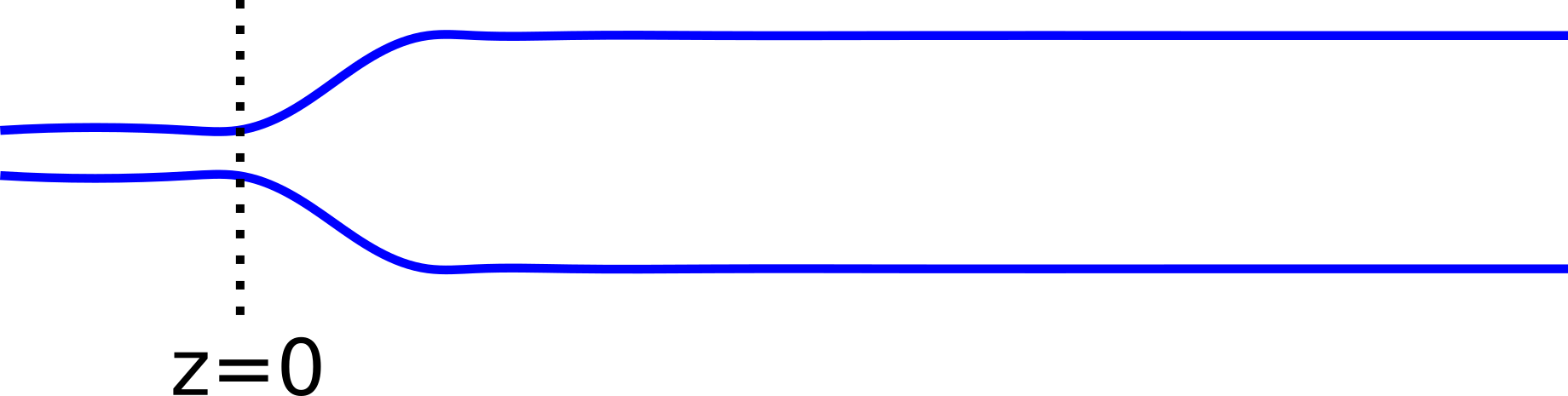}
  \caption{Bolus shape without relaxation}
  \label{fig-without_relax}
\end{subfigure}
\caption{Variation of bolus speed and length before emptying} 
\end{figure*}

To understand how the stiffness and relaxation of the esophageal wall is related to the pressure developed during transport, we have extended FluoroMech by developing a reference model that captures an ideal bolus transport. We define an ideal bolus transport as a bolus moving at a constant velocity without any change in its shape and size. Fig. \ref{fig-reference_model} shows the variation of non-dimensional area with $\chi$ for the FluoroMech reference model. The bolus moves at a constant velocity $u_c$ in the positive $\chi$ direction. This form of area variation can be represented as
\begin{align}
\alpha = f(\chi - u_c\tau)=f(z),	\label{eqn-areaRM}
\end{align}
 where $f$ is some function of $\chi$ and $\tau$. $\chi$ and $\tau$ are combined to form a new transformed coordinate $z=\chi-u_c\tau$. Using Eqs. \ref{eqn-continuity_nd} and \ref{eqn-areaRM}, and converting the spatial and temporal derivatives in terms of derivatives with respect to $z$, i.e. $\partial/\partial \chi = \partial/\partial z$, $\partial/\partial \tau = -u_c\partial/\partial z$, the flow rate $q_r$ is obtained as follows:
\begin{align}
q_r = \int_0^z u_c\frac{\partial \alpha}{\partial z}dz = u_cf(z) + q_{ro},	  \label{eqn-flowRM}
\end{align} 
where $q_{ro} = -u_c$. The subscript $r$ is used to represent quantities in the FluroroMech reference model. It should be noted that the flow rate $q_r$ is calculated between the proximal end of the bolus to the distal end of the esophagus. The flow rate on the proximal side of the bolus can be estimated using Eq. \ref{eqn-flowRM} for negative values of $z$. Assuming the viscous effect to be negligible, we substitute Eq. \ref{eqn-areaRM} and \ref{eqn-flowRM} into Eq. \ref{eqn-momentumRM}, and convert the temporal and spatial derivatives to derivatives in terms of $z$, and obtain the following simplified form:
\begin{align}
\frac{\partial p_r}{\partial z} = \frac{q_{ro}^2}{f^3}\frac{\partial f}{\partial z}.  \label{eqn-P_gradientRM}
\end{align}
Assuming $p_r=0$ at $z=0$, the solution of Eq. \ref{eqn-P_gradientRM} gives the following form for pressure:
\begin{align}
p_r = \frac{u_c^2}{2\alpha^2}\left(\alpha^2-1\right).   \label{eqn-pressureRM}
\end{align}
As reported in \cite{KWIATEK201182}, the fluid pressure developed within the esophagus is directly proportional to the cross-sectional area of the esophageal lumen. Hence, the pressure can estimated using a tube-law of the form:
\begin{align}
p_r = k\left(\alpha-1\right),  \label{eqn-tube_law}
\end{align}
where $k$ is the non-dimensional stiffness of the esophageal wall. The dimensional form of the stiffness can be obtained by multiplying with $\rho c^2$. This measure of stiffness captures the dynamic effect of the esophagus wall since it is calculated using dynamic fluid pressure. The total stiffness will be the combined effect of the static and dynamic intrabolus pressure. From Eqs. \ref{eqn-pressureRM} and \ref{eqn-tube_law}, we obtain the expression for $k$ as follows:
\begin{align}
k = \frac{u_c^2}{2\alpha^2}\left(\alpha+1\right).  \label{eqn-stiffnessRM}
\end{align}
Since $\alpha > 1$, from Eq. \ref{eqn-stiffnessRM} we see that the magnitude of $k$ decreases with increase in $\alpha$. Hence, the minimum stiffness corresponds to the maximum cross-sectional area. This captures the effect of active relaxation of the esophageal wall to accomodate the bolus. Without active relaxation, we would expect $k$ to be independent of $\alpha$. Note that to obtain the expression for $q_r$, $p_r$ and $k$, we integrated from $z = 0$ to any $z > 0$. The point, $z = 0$, always correspond to the proximal end of the bolus, assuming that this point was at $\chi = 0$ at the beginning of the simulation ($\tau = 0$). Therefore, for $k$ to be constant, $\alpha$ should be constant from $z = 0$ to the end of the esophagus. Hence, without relaxation, the bolus would not have such a bulb shape as seen in the FluoroMech reference model in Fig. \ref{fig-with_relax}, and instead would have cylindrical shape of constant diameter that extend from the contraction (at $z = 0$) to the end of the esophagus as shown in Fig. \ref{fig-without_relax}.

From Eq. \ref{eqn-stiffnessRM}, we see that at every point along the length of the esophagus, the stiffness varies from a maximum at $\alpha=1$ to a minimum corresponding to the maximum value of $\alpha$ at that point. Using this fact, we have estimated the active relaxation of the esophageal wall using two different methods. In the first method, the maximum relaxation at a point is approximated as follows: 
\begin{align}
r = \frac{1}{k_{min}}, \label{eqn-relax1}
\end{align}   
where $r$ is the maximum relaxation at a point and $k_{min}$ is the minimum stiffness at the same point. The second method to estimate active relaxation of the esophagus walls is by the use of a relaxation factor $\theta$ such that
\begin{align}
p_r = k_c\left(\frac{\alpha}{\theta} - 1\right), \label{eqn-pressure_relax}
\end{align} 
where $k_c$ is the stiffness of the esophageal wall that remains constant at every $\chi$ and is independent of relaxation. Active relaxation occurs only at the location of the bolus and is not present at other locations where $\alpha=1$. On substituting $\alpha=1$ in Eq. \ref{eqn-stiffnessRM}, we get the constant wall stiffness $k_c = u_c^2$. Using this relation for $k_c$ and Eqs. \ref{eqn-pressureRM} and \ref{eqn-pressure_relax}, we calculate the relaxation factor as follows:
\begin{align}
\theta = \frac{2\alpha^3}{3\alpha^2 - 1}. \label{eqn-relax2}
\end{align}

%%%%%%%%%%%%%%%%%%%%%%%%%%%%%%%%%%%%%%%%%%%%%%%%%%%%%%%%%%%%%%%%%%%%%%%%

\section{Results and Discussions}

\begin{figure*}[h]
 \centering
 \includegraphics[scale = 0.6]{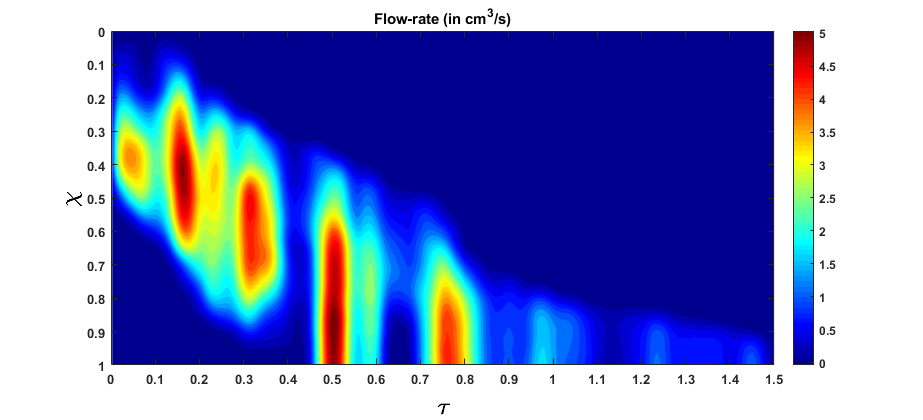} 
 \caption{Flow rate within the esophagus}
 \label{fig-flow_rate}
\end{figure*}

\begin{figure*}[h]
 \centering
 \includegraphics[scale = 0.6]{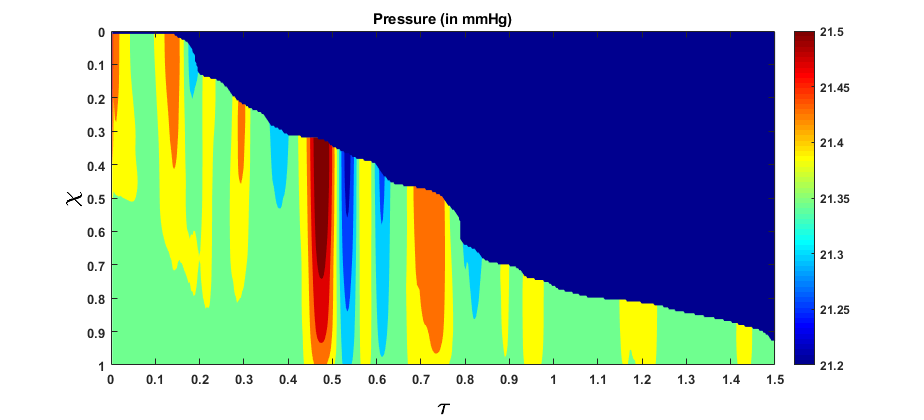} 
 \caption{Fluid pressure within the esophagus}
 \label{fig-pressure}
\end{figure*}

\begin{figure*}[h]
\captionsetup[subfigure]{justification=centering}
\begin{subfigure}[c]{.49\textwidth}
  \centering
  \includegraphics[scale=0.55]{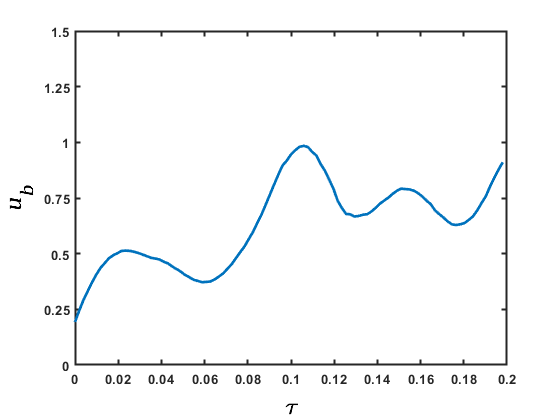}
  \caption{Bulk speed of the bolus}
  \label{fig-ub}
\end{subfigure}
\begin{subfigure}[c]{.49\textwidth}
  \centering
  \includegraphics[scale=0.55]{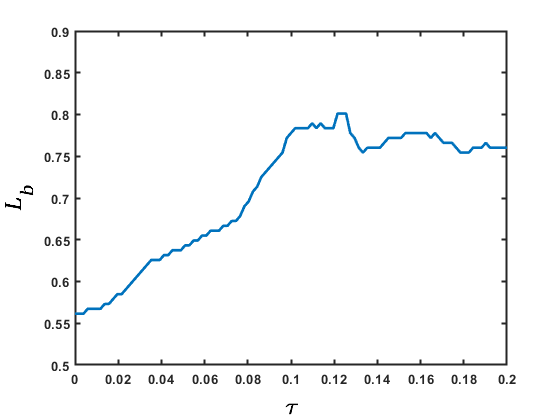}
  \caption{Length of the bolus}
  \label{fig-Lb}
\end{subfigure}
\caption{Variation of bolus speed and length before emptying} 
\end{figure*}

The variation of flow rate and pressure with non-dimensional time and distance along the length of the esophagus is shown in Fig. \ref{fig-flow_rate} and \ref{fig-pressure} respectively. There is no flow at $\chi=1$ for $\tau<0.5$, which indicates pure transport without emptying. The flow rate is non-zero only at the location of the bolus. This matches our observation from the fluoroscopy where the bolus is transported without emptying into the stomach. The variations of area with $\chi$ and $\tau$ lead to fluctuations in pressure (see Fig. \ref{fig-pressure}). As we stated in Section \ref{sec-pressure_variation}, we observe in Fig. \ref{fig-pressure} that the maximum pressure variations can be estimated from the pressure at the proximal end of the bolus. According to Eq. \ref{eqn-pressure_fluc}, these fluctuations in pressure is estimated from the variation of the bulk speed $u_b$ and length $L_b$ of the bolus as shown in Figs. \ref{fig-ub} and \ref{fig-Lb} respectively. Although $p$ is calculated over the whole domain, the fluid pressure within the bolus is the most accurate description of the actual transport process. This is because the VFSE provides information only about the shape of the bolus. Therefore, we have ignored the calculated pressure proximal to the bolus and replaced with a reference value of zero. Additionally, the peristaltic contraction at the proximal end of the bolus, where the diameter is significantly less, cannot be observed in the fluoroscopy, and is fully occluded most of the time, thereby dividing the fluid domain into two parts. 
Our FluoroMech model does not incorporate this contraction zone, and instead treat this zone the same as the remainder of the non-bolus domain. Consequently, we do not see a high-pressure zone at the proximal end of the bolus as seen in the Fig. \ref{fig-ept}. The calculated fluid pressure is the sum of the contributions from the static and dynamic fluid pressures inside the fluid. In this particular senario, the static pressure is the intragastric pressure specified as a boundary condition at the distal end of the esophagus and is equal to 21.4 mmHg. The static pressure is independent of the flow and remains constant throughout the fluid domain. On the other hand, the dynamic pressure depends only on the fluid flow and is shown as the pressure variations in Fig. \ref{fig-pressure}. From the magnitude of the fluid pressure variations, we see that the dynamic pressure is 2 orders of magnitude smaller than the static pressure. The manometer readings of fluid pressure inside the bolus mainly represent the static pressure, and in this case, the measured intrabolus pressure lies between 20 - 25 mmHg. The dynamic pressure variations are too small to be accurately measured by manometry, and so cannot be validated with manometry data. The magnitude of pressure from manometry at the contraction is 50-110 mmHg (shown in Fig. \ref{fig-ept}). Therefore, the contraction pressure is roughly 3 orders of magnitude greater than the dynamic fluid pressure predicted by FluoroMech.

Emptying begins at $\tau=0.5$, which corresponds to $q>0$ at $\chi=1$. This continues untill all fluid is emptied into the stomach. From Fig. \ref{fig-flow_rate} and \ref{fig-pressure} we see that during emptying, a high flow rate corresponds to a high intrabolus pressure. In our model, the reference intragastric pressure is specified at $\chi=1$, so, a high flow rate at $\chi=1$ requires a higher pressure to be developed inside the esophagus to drive the fluid out.  This high pressure inside the bolus indicates the presence of the LES and shows how it behaves differently from the remainder of the esophagus. During normal esophageal transport, the walls distend to accomodate the incoming bolus, and contract back to their relaxed state once the bolus has passed. However, the LES does not distend like the rest of the esophagus, hence a greater pressure inside the bolus is required for the fluid to traverse it. 

\begin{figure*}
\begin{subfigure}[c]{.49\textwidth}
  \centering
  \includegraphics[scale=0.55]{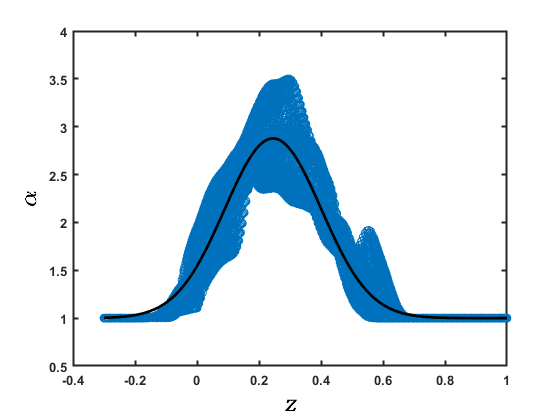}
  \caption{Variation of area with z (for all x and t). The black line shows the fitted gaussian curve}
  \label{fig-bolus_shape}
\end{subfigure}
\begin{subfigure}[c]{.49\textwidth}
  \centering
  \includegraphics[scale=0.55]{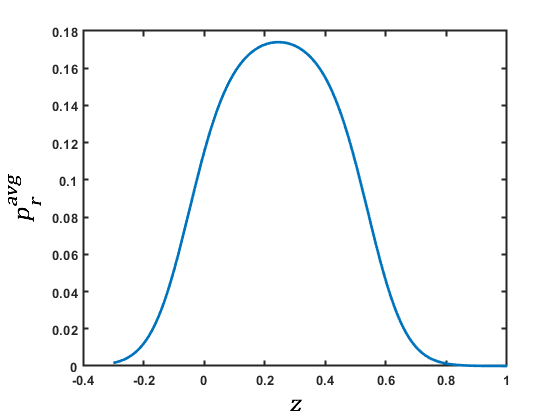}
  \caption{Pressure calculated using the FluoroMech reference model formulation with constant bolus shape}
  \label{fig-pressureRM}
\end{subfigure}
\begin{subfigure}[c]{.49\textwidth}
  \centering
  \includegraphics[scale=0.55]{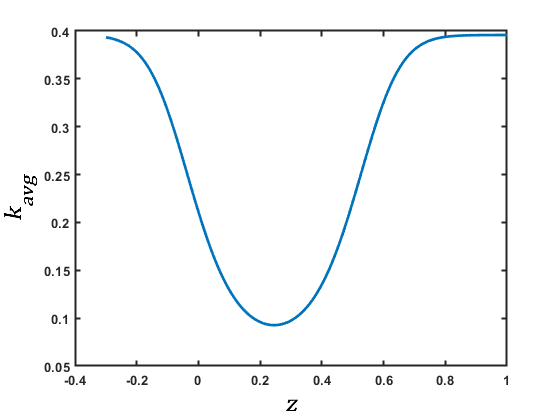}
  \caption{Stiffness calculated using the FluoroMech reference model formulation with constant bolus shape}
  \label{fig-K_avg}
\end{subfigure}
\begin{subfigure}[c]{.49\textwidth}
  \centering
  \includegraphics[scale=0.22]{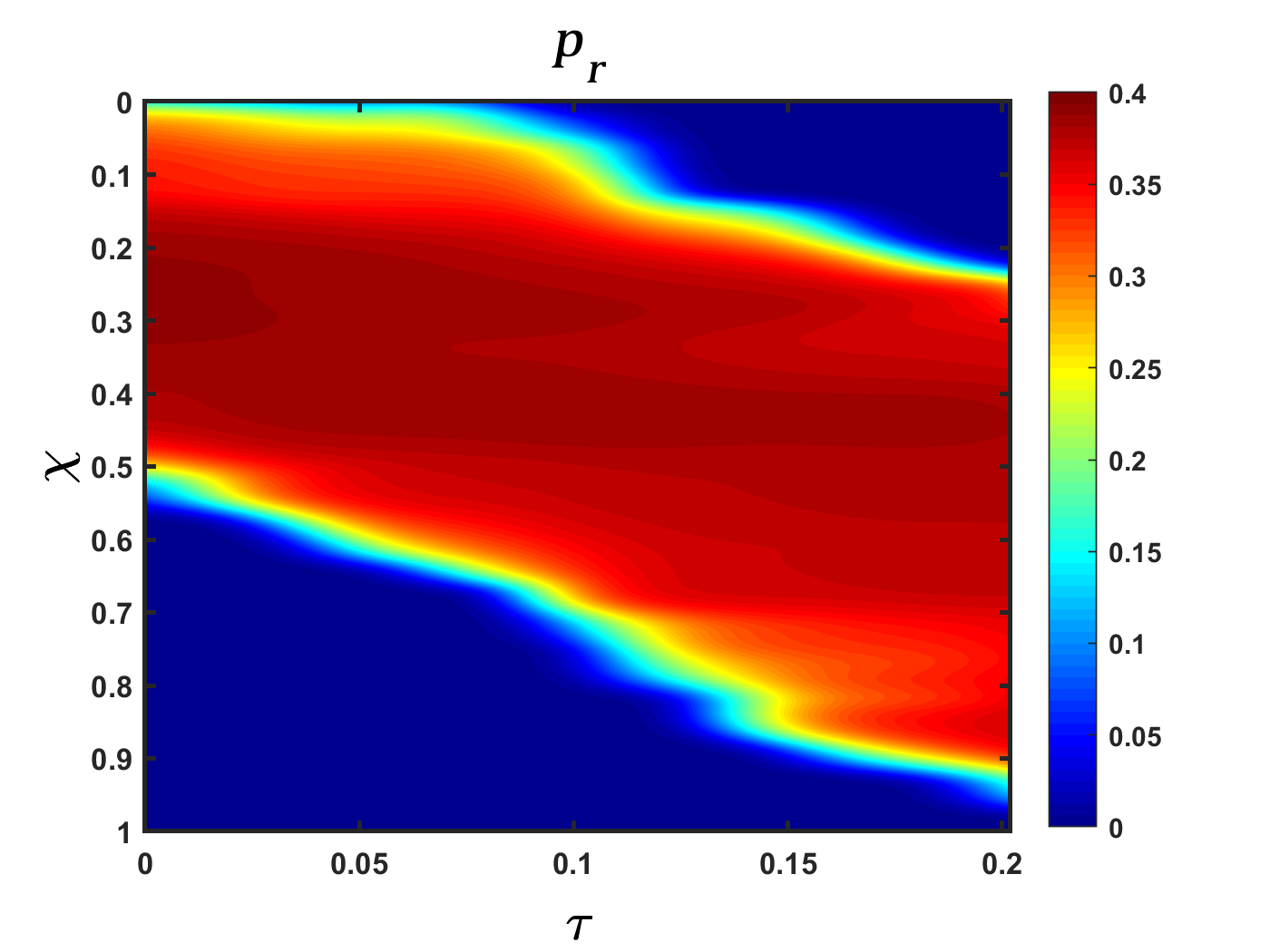}
  \caption{Pressure calculated using the FluoroMech reference model formulation with varying bolus shape}
  \label{fig-p_r}
\end{subfigure}
\begin{subfigure}[c]{.49\textwidth}
  \centering
  \includegraphics[scale=0.55]{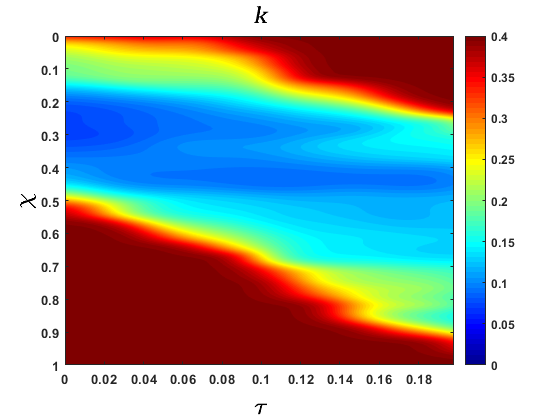}
  \caption{Stiffness calculated using the FluoroMech reference model formulation with varying bolus shape}
  \label{fig-K}
\end{subfigure}
\begin{subfigure}[c]{.49\textwidth}
  \centering
  \includegraphics[scale=0.55]{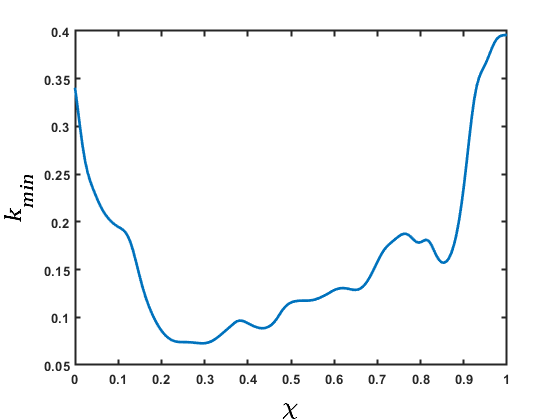}
  \caption{Minimum stiffness at every location along the length of the esophagus}
  \label{fig-K_min}
\end{subfigure}
\caption{Estimation of esophageal stiffness}
\end{figure*}
We have extended our analysis to estimate the stiffness of the esophageal wall and the active wall relaxation that occurs ahead of the peristaltic contraction in order to accomodate the incoming bolus. This is done by utilizing the FluoroMech reference model described in Section \ref{sec-stiffness}. The main assumptions for using the reference model formulation were a) that the bolus moves at a constant velocity, b) that bolus shape does not change, and c) that the viscous effect is negligible. To satisfy assumption (a) for the reference model, the constant bolus velocity $u_c$ is estimated by taking the mean of the bulk bolus velocity $u_b$ (as shown in Fig. \ref{eqn-u_b}) averaged over time. In order to satisfy the assumption (b), we fitted a single Gaussian curve of the form $\alpha = a\textrm{exp}[-((z-b)/d)^2]+\alpha_o$ on the scatter plot of the non-dimensional cross-sectional area of the bolus with respect to the transformed coordinate $z$ shown in Fig. \ref{fig-bolus_shape}. This fitted curve provides an estimate of the constant shape of the bolus in the reference model. For the given case, $a = 1.877$, $b = 0.245$, $d = 0.220$, and $\alpha_o=1$. Using the properties of water for the swallowed fluid, we have observed that the viscous term of Eq. \ref{eqn-momentum} is very small compared to each of the known inertia terms as well as the sum of the inertia terms. Hence, the viscous term was dropped to calculate pressure without a significant error, thus satisfying assumption (c) of the reference model. 

Using the constant speed $u_c$ and the constant shape of the bolus, we calculated an average pressure and average stiffness according to Eqs. \ref{eqn-pressureRM} and \ref{eqn-stiffnessRM} respectively. Figs. \ref{fig-pressureRM} and \ref{fig-K_avg} show the pressure and stiffness calculated using the FluoroMech reference model formulation. Therefore, the fluctuations of pressure (see Fig. \ref{fig-pressure}) that do not match the tube-law pressure form according to Eq. \ref{eqn-tube_law} are eliminated. On comparing the maximum values of non-dimensional $p$ and $p_r^{avg}$, we have seen that the dynamic variations of $p$ are 2 orders of magnitude greater than $p_r^{avg}$. Hence, the fluctuations of $p$ completely dominate the tube-law form of pressure. It is difficult to identify the sources of these fluctuations from the two-dimensional fluoroscopy study. These fluctuations may be due to the esophagus not being completely horizontal, and gravity effects causing the bolus to accelerate or decelerate depending on the esophageal orientation as a whole as well as the irregularities in the mucosal surface of the lumen. The esophagus might deform as well due to contact with surrounding organs, which in turn vary with time due to the heart beating, vasculature pulsating, respiration, and overall body movement leading to variations in cross-sectional areas and consequently the pressure. The FluoroMech reference model eliminates these fluctuations and captures the essence of bolus transport giving a measure of pressure and stiffness that is consistent with the tube-law. From Fig.\ref{fig-K_avg}, we see that the average stiffness $k_{avg}$ is minimum for the corresponding maximum cross-sectional area due to the relaxation of the esophageal wall to accomodate an incoming bolus.

In order to calculate the variations of pressure and stiffness with $\chi$ and $\tau$, these quantities are calculated using the actual area instead of the approximate area of the bolus. The actual area is substituted in Eqs. \ref{eqn-pressureRM} and \ref{eqn-stiffnessRM} to calculate pressure and stiffness respectively. The pressure and stiffness variations over $\chi$ and $\tau$ are shown in Figs. \ref{fig-p_r} and \ref{fig-K} respectively. 

\begin{figure*}[h]
  \centering
  \includegraphics[scale=0.7]{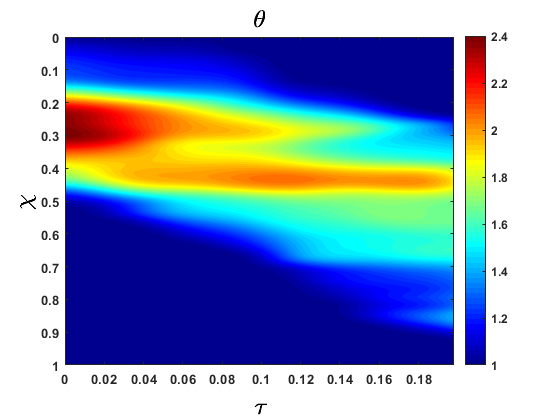}
\caption{Esophageal active relaxation estimated by relaxation factor}
\label{fig-relax2}
\end{figure*}
We presented two methods for estimating the relaxation of the esophagus wall. In the first method, described by Eq. \ref{eqn-relax1}, the minimum stiffness at each location along the length of the esophagus (as shown in Fig. \ref{fig-K_min}) is used to approximate active wall relaxation. In the second method, the active relaxation of esophageal wall is estimated using a relaxation factor $\theta$ according to Eq. \ref{eqn-relax2}, the variation of which is shown in Fig. \ref{fig-relax2}. From Figs. \ref{fig-K_min} and \ref{fig-relax2}, we see that the maximal esophageal relaxation occurs at the same location, i.e. at $\chi=0.3$, and is quantified to have a value of $16$ and $4.9$ using the two methods according to Eqs. \ref{eqn-relax1} and \ref{eqn-relax2} respectively. These estimates of relaxation are very sensitive to the relaxed cross-sectional area $A_o$ of the esophageal lumen. Identifying the lumen in the fluoroscopy images is very difficult for the given case, and further information on $A_o$ through computed tomography (CT) or magnetic resonance (MR) imaging would be more accurate in estimating esophageal active wall relaxation.

%%%%%%%%%%%%%%%%%%%%%%%%%%%%%%%%%%%%%%%%%%%%%%%%

\section{Conclusion}

We have presented FluoroMech, a technique for analyzing fluoroscopy image data using deep learning and computational fluid dynamics. The image sequence from fluoroscopy was segemented using a Convolutional Neural Network to obtain the outline of the bolus as it transits the esophagus. This bolus outline then becomes the input to a computational model that solves the one-dimensional mass and momentum conservation equations to obtain the fluid flow rate and pressure. Since fluoroscopy provides information only about the shape of the bolus in a single two-dimensional plane, we made approximations regarding the shape of the cross-section in order to conserve the volume of swallowed fluid. Our model indicates the LES behaves very differently from the rest of the esophagus, in that it acts as a restriction to the outflow of fluid from the esophagus by not expanding as easily as the remainder of the esophagus. Thus, FluoroMech approximately quantifies the behavior of the LES in terms of the pressure difference between the intrabolus and intragastric pressure, and flow rate at the distal end of the esophagus. 

Using the shape of the bolus and the velocity predicted from this model, we have presented a method, called the FluoroMech reference model, to estimate esophageal wall stiffness and active relaxation. The reference model eliminates the undesirable variations in cross-sectional area that lead to pressure fluctuations, and uses the relevant pressure predictions to capture the state and functioning of the esophagus.  

Based on our mechanistic study, we have categorized the esophageal transport into four zones: contraction zone behind the bolus, active relaxation zone at the bolus, stiff zone at the LES and a baseline zone for the remainder of the esophagus.

\bigbreak

\textbf{Acknowledgements} We acknowledge the support provided by Public Health Service Grants R01-DK079902 and P01-DK117824, and National Science Foundation Grants OAC 1450374 and OAC 1931372 in the completion of this work. We also acknowledge the computational resources provided by Northwestern University's Quest High Performance Computing Cluster. For this work, we have also utilized the Extreme Science and Engineering Discovery Environment (XSEDE) cluster Comet, at the San Diego Supercomputer Center through allocation TG-ASC170023, which is supported by National Science Foundation grant number ACI-1548562 \cite{xsede}.

\bibliographystyle{plain}
\bibliography{main}

\end{document}